\documentclass[aps, prc, 11pt, preprintnumbers,amsmath,amssymb]{revtex4}
\usepackage{amsmath}
\usepackage{amssymb}
\usepackage{color}
\usepackage{mathrsfs}
\usepackage{graphics}
\usepackage{graphicx}
\usepackage{dcolumn}
\usepackage{bm}
\def\ni{\noindent}
\begin{document}
\title{A Tale of Two Mergers:  Searching for Strangeness in Compact Stars } 

\author{Madappa Prakash}
\email{prakash@snare.physics.sunysb.edu}

\author{James M. Lattimer}
\email{lattimer@pulsar.ess.sunysb.edu}

\affiliation{Department of Physics \& Astronomy \\ 
        State University of New York at Stony Brook \\
        Stony Brook, NY 11794-3800, USA}
\begin{abstract}

We contrast the evolution of gravitational binary mergers for the two
cases of a black hole and a normal neutron star, and a black hole and
a self-bound strange quark matter star.  In both cases inspiral
continues until the Roche limit is reached, at which point it is
expected that stable mass transfer to the balck hole ensues.  Whereas
a neutron star would then outspiral, the strange quark matter star
barely moves out at all.  Eventually, the strange quark star loses all
its mass to the black hole, as compared to the neutron star whose
outspiral continues until stable mass transfer terminates and inspiral
resumes.  These scenarios result in distinctly different gravitational
wave signatures.
\end{abstract}
\maketitle

%%%%%%%%%%%%%%%%%%%%%%%%%%%%%%%%%%%%%%%%%%%%%%%%%%%%

\section{Introduction}
\label{sec:Introduction}

\begin{figure}[hbt]
\begin{center}
\includegraphics[scale=0.55,angle=0]{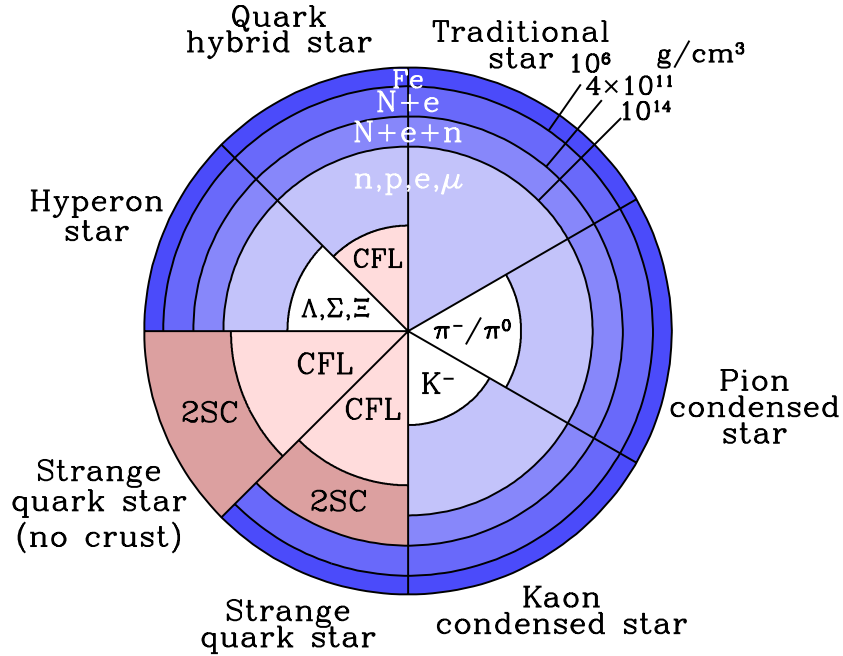}
\end{center}
\vspace*{-0.5in}
\caption[]{A schematic diagram of the possible phases in a  
compact star (Courtesy: A.W. Steiner). }  
\label{nstruct}
\end{figure}

Several proposals concerning the physical state and the internal
constitution of matter at supra-nuclear densities have been put forth
(see~\cite{LP01,Alford01} for recent accounts).  Figure~\ref{nstruct} shows
many exciting possibilities for the composition of compact stars
including (1) strangeness-bearing matter in the form of hyperons,
kaons, or quarks, (2) Bose (pion or kaon) condensed matter, and (3)
so-called self-bound strange quark matter (SQM).  Fermions, whether
they are in the form of baryons or deconfined quarks, are expected to
additionally exhibit superfluidity and/or superconductivity. In the
case of quarks, this could occur either as two-flavor
superconductivity (2SC) or in the Color-Flavor-Locked (CFL) phase in
which all three flavors (up, down, and strange)
participate.

%\clearpage

The possibility of such exotic phases brings attendant changes to the
predictions of maximum masses and radii, since the presence of
multiple components or new phases of matter generally lessens the
pressure for a given energy density. 
The fact that compact objects are the only objects in which such
phases could occur underscores the importance of precise
determinations of basic observables:

\ni (i) masses and radii~\cite{LP01}. Simply put, a precise
determination of the mass and radius of the same neutron star would not
only be a first achievement for observational astronomy, but would also hold
the promise of delineating Quantum Chromo-Dynamics (QCD) in regimes of high
baryon density heretofore not possible. 

\ni (ii) surface temperatures versus age~\cite{PPLS00}.  Detections of
photons from cooling neutron stars could constrain the pairing gaps
of baryons and the star's mass. However, establishing the sizes of quark
gaps or the presence of Goldstone bosons in the CFL phase of quarks
in a hybrid star 
will be difficult~\cite{PPLS00,JPS02} because the layers of
normal matter surrounding the quark core continues to cool
through significantly more rapid processes.  In contrast, the
mean energy of emitted photons from the bare surface of a strange
quark star would be significantly larger than that from a normal
cooling neutron star ($30 <E/{\rm keV} < 500$ versus $0.1 < E/{\rm
keV} <2.5$)~\cite{Page02b}.  Due to its distinctive spectrum and time
evolution, such an observation would constitute an unmistakable
detection of a strange quark star and shed light on color
superconductivity at ``stellar'' densities. 

\ni (iii) anomalous behavior in the spin-down rates of ms 
radio pulsars~\cite{GPW97}. Possible hysteresis (reversal) in the
normal spindown rate of pulsars would signal the
appearance of a mixed-phase of matter containing normal hadronic
matter with more compressible Bose condensed or quark matter.

\ni (iv) neutrino luminosities from future galactic core
collapse supernovae. The main new feature that has emerged from
studies of neutron stars at birth is the possible metastability and
subsequent collapse to a black hole of a proto-neutron star containing
quark matter, or other types of matter including hyperons or a Bose
condensate, which could be observable in the neutrino
signal~\cite{Pons01b}.

Examples of ongoing and planned observations of solar mass compact
objects that could shed light on QCD at high baryon density  include

\ni (i) multi-wavelength photon observations with the HST, Chandra,
XMM, Integral, {\em etc}., 

\ni (ii) spectral and temporal studies of supernova neutrinos with
the SuperK, SNO, UNO, {\em etc}., and 

\ni (iii) gravity wave detections from coalescing compact object binaries
though LIGO, VIRGO,  {\em etc}.

In this work, we contrast the evolution of binary star mergers for two
distinct cases: 

\ni (1) A black hole (BH) and a normal star.  For the discussion
at hand, this refers to a star with a surface of normal matter in
which the pressure vanishes at vanishing baryon density.  The interior
of the star, however, may contain any or a combination of the many 
exotica~{\footnote{Not to be confused with the Canadian flick 
``Exotica.''}} depicted in Fig.~\ref{nstruct}.

\ni (2) A BH and a self-bound star which is exemplified by Witten's
Strange-Quark Matter (SQM) star~\cite{Witten}; see~\cite{Alcock} for a
review.  Such a star has a bare quark matter surface in which the pressure
vanishes at a finite but supra-nuclear baryon density~{\footnote{In
the context of the MIT bag model with first order corrections due to
gluon exhange, the baryon density at which pressure vanishes is given
by $n_b(P=0) = (4B/3\pi^{2/3})^{3/4}(1-2\alpha_c/\pi)^{1/4}$, where
$B$ is the bag constant and $\alpha_c=g_c^2/(4\pi)$ is the quark-gluon
coupling constant~\cite{PBP90}. This density is not significantly
affected by the finite strange quark mass~\cite{PBP90} or by pairing
gaps in the CFL or 2SC phases~\cite{JMad}.}}.

\begin{figure}[hbt]
\includegraphics[width=0.45\textwidth]{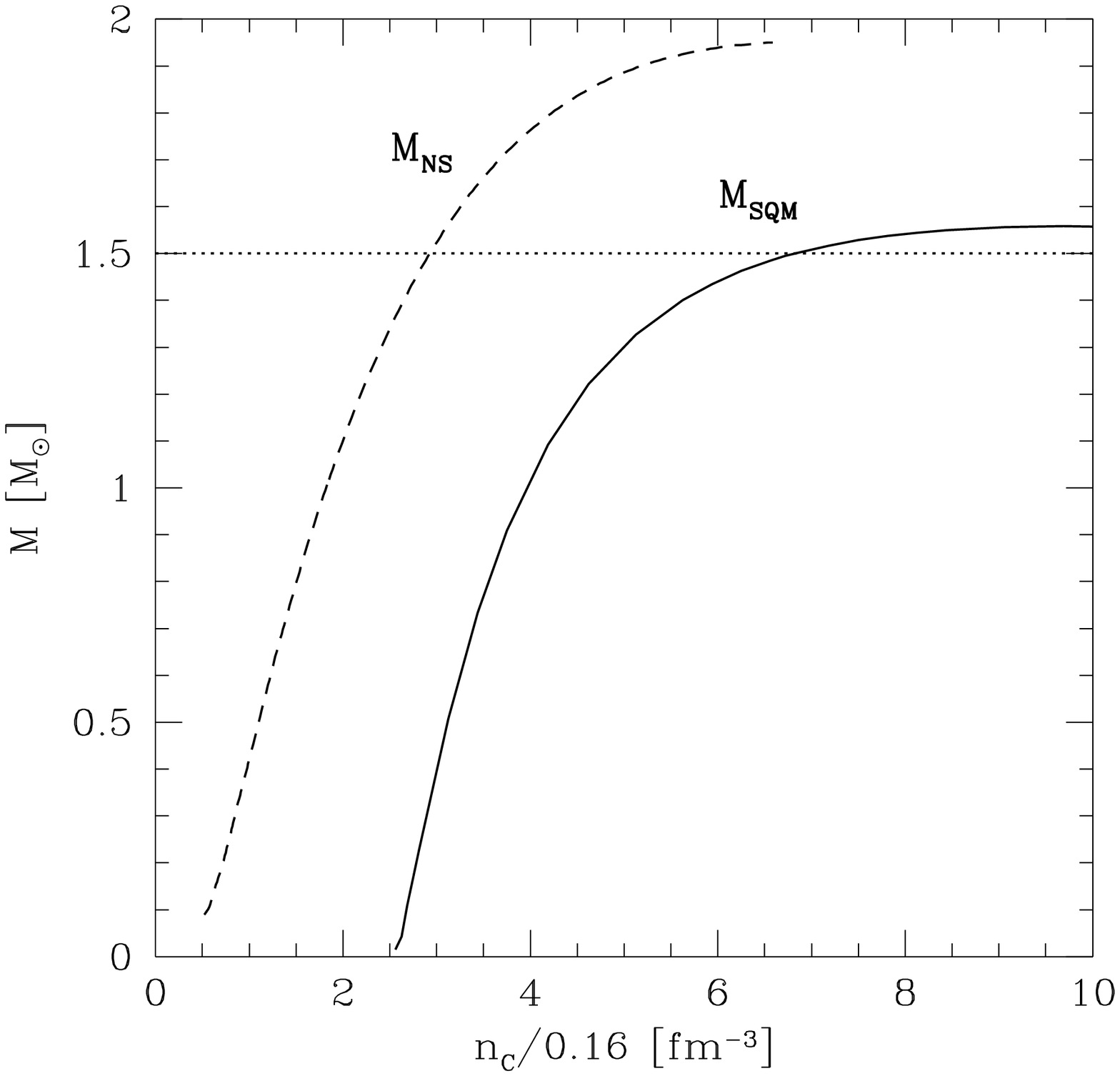}
\includegraphics[width=0.45\textwidth]{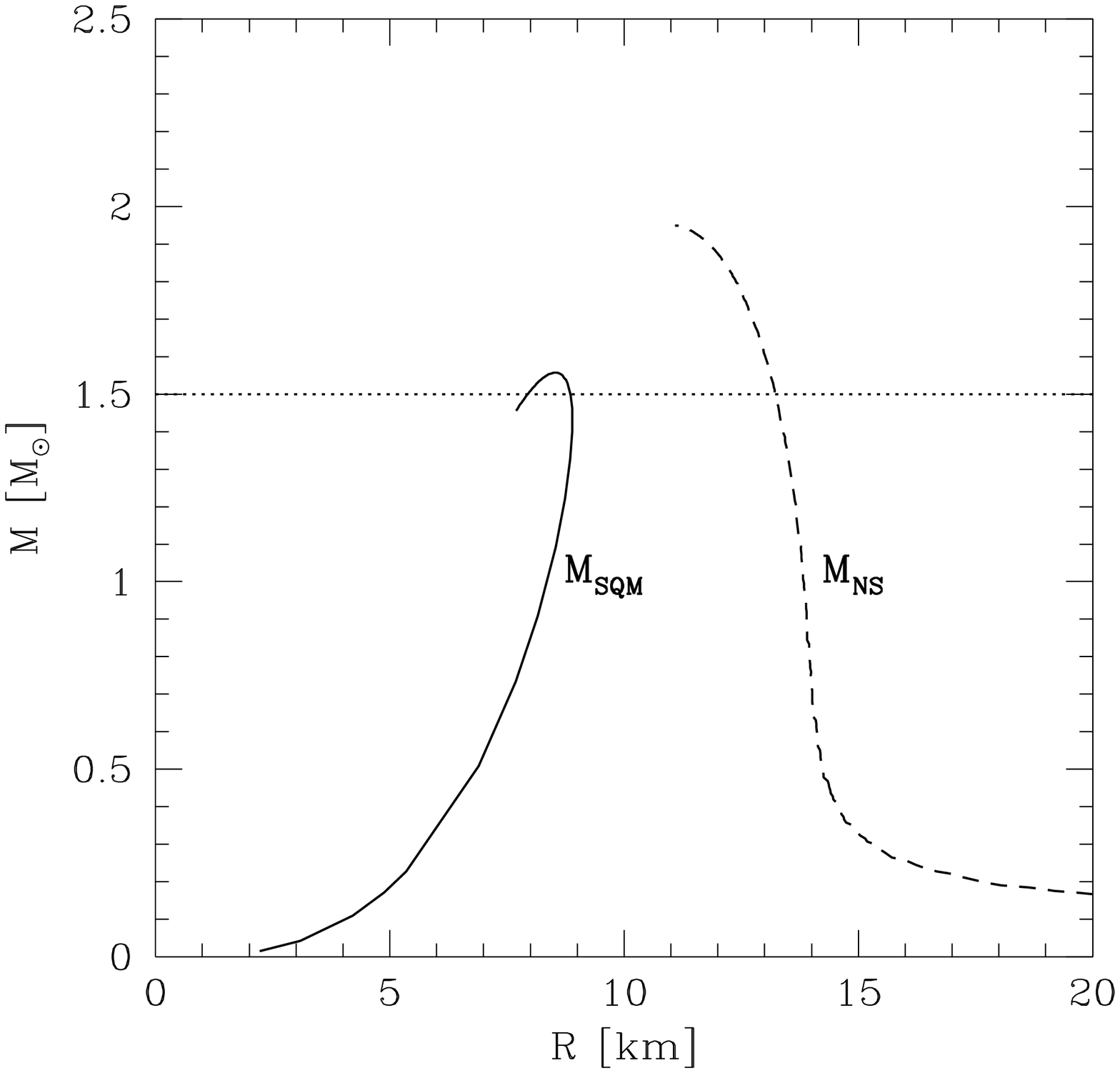}
\vspace*{-0.5cm}
\caption{(a) Mass versus central density, and (b) mass versus radius,
for normal and self-bound stars. }
\label{mnr}
\end{figure}

Prototypes of the mass versus central baryon density $(n_c)$ and mass
versus radius for these cases are shown in
Fig.~\ref{mnr}. Quantitative variations from these generic behaviors
can be caused by uncertainties in the underlying strong interaction
models (see the compendium of results in Fig. 2 of
Ref.~\cite{LP01}). Qualitative differences in the outcomes of
mergers with a black hole emerge, however, because of the gross
differences in the mass-radius diagram.

\begin{figure}[hbt]
\includegraphics[width=0.55\textwidth]{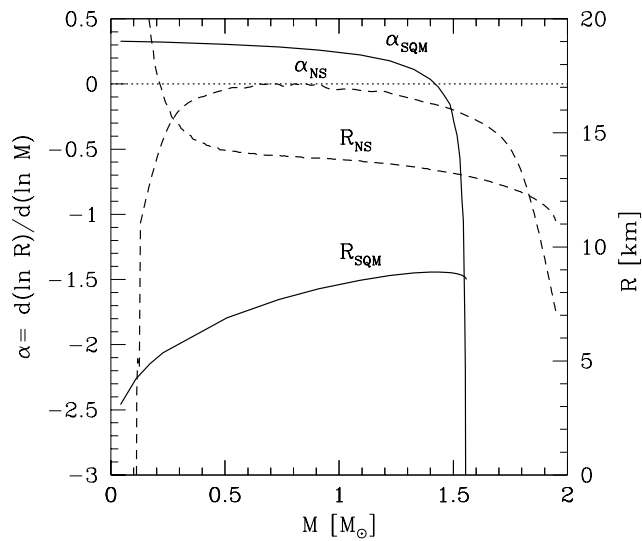}
\vspace*{-0.5cm}
\caption{Comparisons of $\alpha$ and radii $R$ for the normal star (NS) and
self-bound strange quark matter (SQM) star as functions of mass.  }
\label{alpha}
\end{figure}

A normal star and a self-bound star represent two quite different 
possibilities (see Fig.~\ref{alpha}) for the quantity
\begin{eqnarray}
\alpha \equiv  \frac {d\ln R}{d\ln M} 
\left\{ \begin{array}{ll} 
\leq 0 & \mbox{{\rm for~a~normal~neutron~star~(NS)}} \\
\geq 0 & \mbox{{\rm for~a~self-bound~SQM~star}} 
\end{array} 
\right. \,, 
\label{lderiv}
\end{eqnarray}
where $M$ and $R$ are the star's mass and radius, respectively. For
small to moderate mass self-bound stars, $R \propto M^{1/3}$ so that
$\alpha \cong 1/3$; only for configurations approaching the maximum mass
does $\alpha$ turn negative.

%

%\section{Motivation} 

Our objective is to explore the astrophysical consequences of these
distinctive behaviors in $R$ versus $M$ as they affect mergers with a
black hole.  Note that $\alpha$ is intimately connected with the dense
matter equation of state (EOS), since there exists a one-to-one
correspondence between $R(M)$ and $P(n_B)$, where $P$ is the pressure
and $n_B$ is the baryon density.  
Gravitational mergers in which a compact star loses its mass (either
to a companion star or to an accretion disk) during evolution is one
of the rare examples in which the $R$ versus $M$ (or equivalently, $P$
versus $n_B$) relationship of the same star is sampled.
Although we focus here on the coalescence of a compact star
with a BH, the theoretical formalism and our principal findings apply
also to mergers in which both objects are compact stars.

%xxxxxxxxxxxxxxxxxxxxxx

\section{The Merger of a Compact Star with a Low-Mass Black Hole}

The general problem of the origin and evolution of systems containing
a neutron star and a black hole was first detailed by Lattimer \&
Schramm~\cite{LSch,thesisL}. Compact binaries form naturally as the
result of evolution of massive stellar binaries ({\em e.g.,} ~\cite{BB}).  
The estimated lower
mass limit for supernovae (which produce neutron stars or black holes)
is approximately 8 M$_\odot$.  Observationally, the number of binaries
formed within a given logarithmic separation is approximately
constant, but the relative mass distributions are uncertain.  There is
some indication that the distribution in binary mass ratios might also
be flat.  Most progenitor systems do not survive the first, more massive
star becoming a supernova. In the absence of a ``kick'' velocity from
the explosion, it is
easily found that the loss of more than half of the mass from the
system will unbind it.  However, the fact that pulsars are observed to
have mean velocites in excess of a few hundred km/s implies that
neutron stars are usually produced with large kicks.
In the cases that the kick, which is thought to be randomly directed, opposes the star's
orbital velocity, the chances that the post-supernova binary remains
intact increases.  In addition, the orbital separation in a surviving binary
will be reduced significantly.  Subsequent evolution then progresses
to the supernova explosion of the companion.  Those systems that
survive the second explosion should both have greatly reduced separations and orbits with high
eccentricity.

Gravitational radiation causes the binary's orbit to decay~\cite{HT}, such
that a system with masses $M_1$ and $M_2$ with initial
semimajor axes $a$ satisfying
\begin{equation}
a<2.8[M_1M_2(M_1+M_2)/{\rm M}_\odot^3]^{1/4} {\rm R}_\odot\,,
\label{decay}
\end{equation}
will fully decay within the age of the Universe ($\sim10^{10}$ yr).
This limit is for circular orbits; those with highly eccentric orbits will decay much faster~\cite{Peters}.
Ref.~\cite{LSch} argued that (1) mergers of neutron stars and black
holes coupled with the subsequent ejection of a few percent of the neutron
star's mass, could easily account for all the {\em r}-process nuclei
in the cosmos, and (2) compact object binary mergers could be associated
with gamma-ray bursts (see also~\cite{eichler}).

\begin{figure}[hbt]
\includegraphics[width=0.55\textwidth]{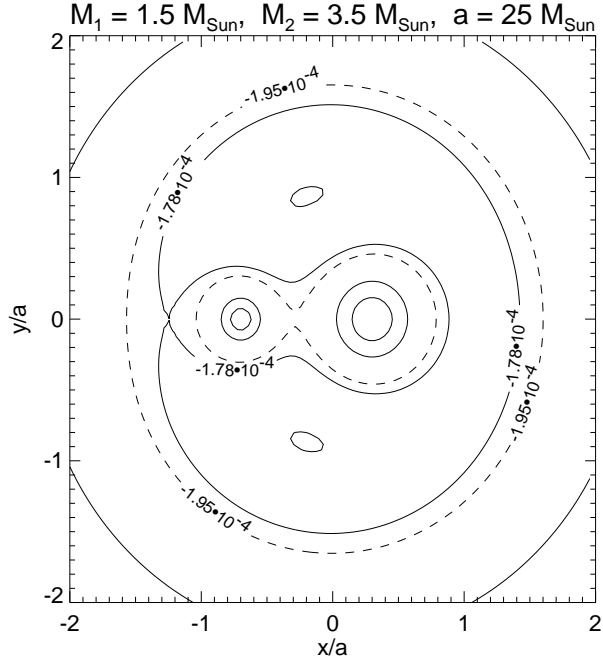}
\vspace*{-0.5cm}
\caption{Equipotential contours of Newtonian gravitational plus
centrifugal potential of a compact star and a black hole binary. The
inner dashed curve shows the Roche lobe (Courtesy: S. Ratkovic).}
\label{rlobe}
\end{figure}

When the less massive inspiralling compact star reaches its Roche limit (see
Fig.~\ref{rlobe}), mass can be stripped from it. The radius of the
compact object will quickly adjust to its new mass. If the radius
increases more quickly than the Roche limiting radius, mass
transfer to the BH will be stable, and the inspiral will be halted due
to angular momentum conservation.
The classical Roche limit is based upon an incompressible fluid of
density $\rho$ and mass $M_2$ in orbit about a mass $M_1$.  In
Newtonian gravity, this limit is
\begin{equation}
R_{Roche, Newt}=(M_1/0.0901\pi\rho)^{1/3}= 19.2
(M_1/{\rm M}_\odot
\rho_{15})^{1/3}{\rm~km}\,,\label{roche}
\end{equation}
where $\rho_{15}=\rho/10^{15}$ g cm$^{-3}$.  Using general relativity,
Fishbone~\cite{Fishbone} found that the number 0.0901 in
Eq.~(\ref{roche}) becomes 0.0664, even for rotating BHs.  In
geometrized units, $R_{Roche}/M_1=13(14.4)(M_1^2\rho_{15}/{\rm
M}_\odot^2)^{-1/3}$, where the numerical coefficient refers to the
Newtonian (GR) case.  In other words, if the neutron star's mean
density is $\rho_{15}=1$, the Roche limit is encountered beyond the
last stable orbit ($R=6M_1$ for a non-rotating BH) if $M_1<5.9$
M$_\odot$, leading to mass overflow and
mass transfer.  And, as now discussed, the mass
transfer may proceed stably for some considerable time.  This would
lengthen the lifetime the BH would accrete matter from its companion,
which, in itself, could be an observational signature.

\section{EVOLUTION OF MERGERS} 

\begin{figure}[hbt]
\includegraphics[width=0.475\textwidth,angle=90]{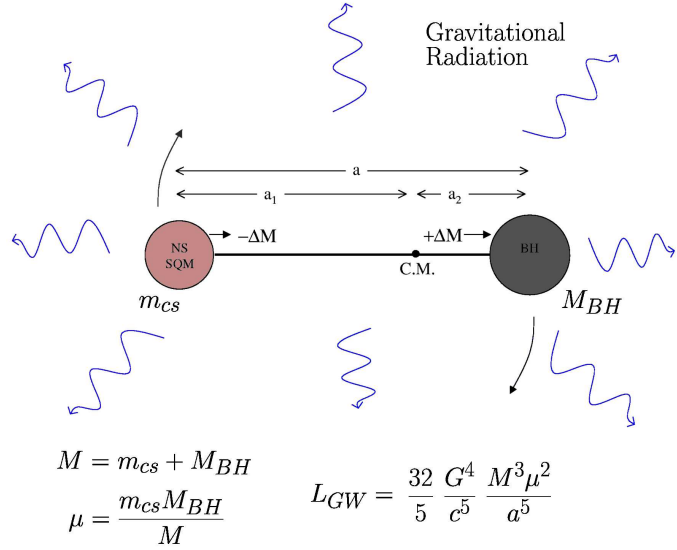}
\vspace*{-0.5cm}
\caption{A schematic illustration of a compact star merger with a
black hole (Courtesy: M.W. Carmell).}
\label{binup}
\end{figure}

The final evolution of a compact binary is now discussed (see
Fig.~\ref{binup} for a schematic illustration).  Define
$q=m_{cs}/M_{BH}$, $\mu=m_{cs}M_{BH}/M$, and $M=M_{BH}+m_{cs}$, where
$m_{cs}$ and $M_{BH}$ are the compact star (normal star or self-bound
SQM star) and black hole masses, respectively. The orbital angular
momentum is
\begin{equation}
J^2=G\mu^2 Ma=GM^3aq^2/(1+q)^4\,.
\label{j2}
\end{equation}
We can employ Paczy\'nski's~\cite{pacz} formula for the Roche radius of the
secondary:
\begin{equation}
R_\ell/a=0.46[q/(1+q)]^{1/3}\,,
\label{rl}
\end{equation}
or a better fit by Eggleton~\cite{eggleton}:
\begin{equation}
R_\ell/a=0.49[0.6+q^{-2/3}\ln(1+q^{1/3})]^{-1}\,.
\label{eggleton}
\end{equation}
The orbital separation $a$ during mass transfer is obtained by setting
$R_\ell=R$, the compact star radius.  For stable mass transfer, the
star's radius has to increase more quickly than the Roche radius as
mass is transferred~{\footnote{Mass extraction from a self-bound SQM
star has been estimated by Madsen~\cite{JMad} by requiring that the
gravitational tidal force exceeds that due to surface tension, {\em
i.e.,} $ G m_{cs}M_{BH} R/a^3 > \sigma R$, where $\sigma$ is the
surface tension of SQM.  This leads to \\ $A > \sigma a^3 /(GM_{BH}) =
4\times10^{38} ~(\sigma/20~{\rm MeV~fm^{-2}}) ~(a/30~{\rm km})^3 ~({\rm
M}_\odot/M_{BH})$, \\ upon using $m_{cs}=Am_u$, where $m_u$ is the
atomic mass.}}.  Thus, we must have, using Paczy\'nski's formula,
\begin{equation}
{d\ln R\over d\ln m_{cs}}\equiv\alpha\ge{d\ln R_\ell\over d\ln m_{cs}}= {d\ln
a\over d\ln m_{cs}}+{1\over3}\label{dlnrdlnm2}
\end{equation}
for stable mass transfer.  If the mass transfer
is conservative~{\footnote{Some aspects of non-conservative mass
transfer are discussed in \cite{LPsch,PZ} and references therein.}}, 
$\dot J=\dot J_{GW}$, where
\begin{equation}
\dot J_{GW}=-{32\over5}{G^{7/2}\over c^5}{\mu^2 M^{5/2}\over a^{7/2}}=
-{32\over5}{G^{7/2}\over c^5}{q^2 M^{9/2}\over(1+q)^4a^{7/2}}\label{dotgw}
\end{equation}
and
\begin{equation}
{\dot J\over J}={\dot a\over2a}+{\dot q(1-q)\over
q(1+q)}\,,\label{dotj}
\end{equation}
where the dots on various symbols denote time derivatives.
This leads to
\begin{equation}
\dot q\left({\alpha\over2}+{5\over6}-q\right)\ge -{32\over5}{G^3\over c^5}{q^2
M^3\over(1+q)a^4}\,.\label{dotq}
\end{equation} 
Since $m_{cs}<M_{BH}$, $\dot q\le0$, and the condition for stable mass
transfer is simply
\begin{equation}
q\le5/6+\alpha/2\,. 
\label{cond}
\end{equation}
This condition is achieved in all normal neutron star - BH or
self-bound star - BH binaries. In both cases, the luminosities of
gravity wave emission can be estimated using
\begin{equation}
L_{GW} = \frac {32}{5} \frac{G^4}{c^5} \frac{M^3\mu^2}{a^5} \,.
\label{lgw}
\end{equation}

\section{RESULTS AND DISCUSSION}

\begin{figure}[hbt]
\includegraphics[width=0.475\textwidth]{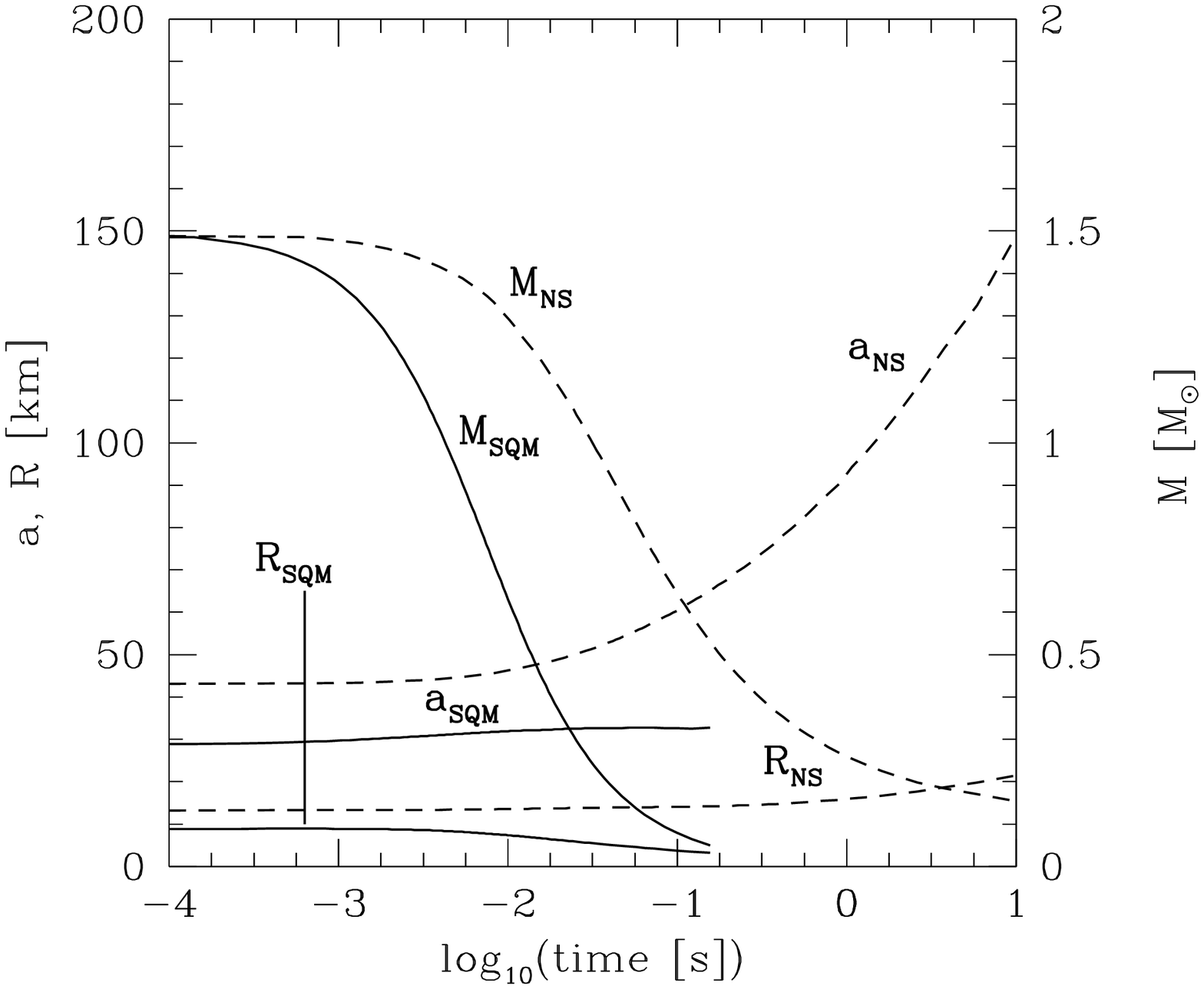}
\includegraphics[width=0.475\textwidth]{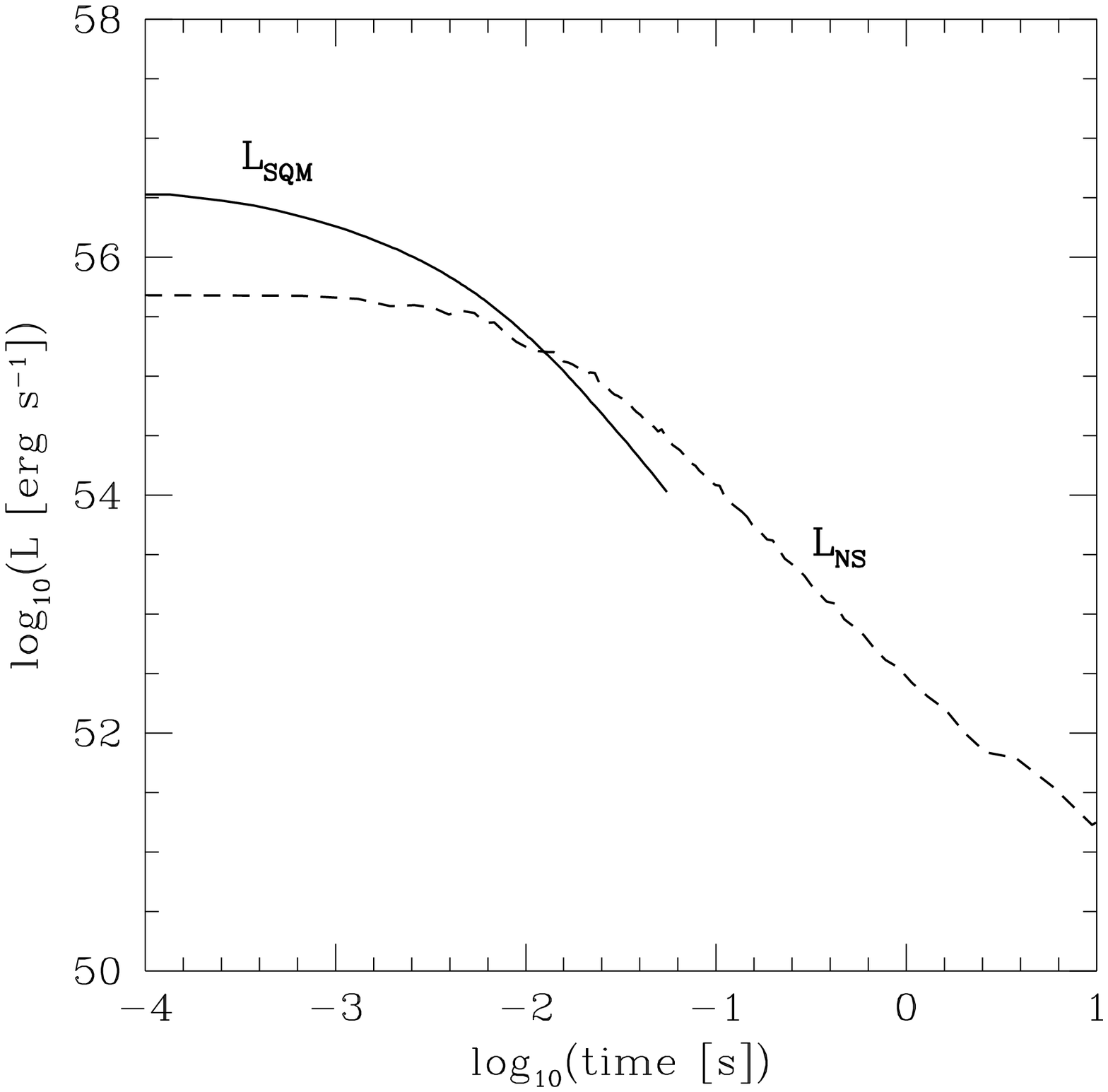}
\vspace*{-0.5cm}
\caption{{\em Left panel}: The evolution of a 1.5 M$_\odot$ 
SQM star (solid line) is compared with that of a 1.5 M$_\odot$ normal
neutron star (dashed line) with a 3.5 M$_\odot$ black hole during
merger.  The compact stars' masses, radii, and outspiral orbital separations
during stable mass transfer are shown. (For simplicity, the inspiral
evolutions are not indicated.)  {\em Right panel}: Luminosities of
gravity wave emission for the two cases.}
\label{orbit}
\end{figure}

In addition to Eq.~(\ref{cond}), a number of other conditions must
hold for stable mass transfer to occur ({\em cf.}~\cite{LPsch}, and
references therein).  The discussion in~\cite{LPsch} is relevant to
the stellar merger of a self-bound star as well.  We now compare
the mergers of normal and self-bound stars with a BH.  For
definiteness, we choose $m_{ns}=m_{SQM}=1.5{\rm M}_\odot$ and
$M_{BH}=3.5{\rm M}_\odot$. Results for other choices are
straightforward to obtain.  Fig.~\ref{orbit} (left panel) shows the
time development of the orbital separation $a$ and the compact star's
mass and radius during the stable mass transfer (outspiral) phases.
The time evolutions during stable mass transfer are obtained from
Eq.~(\ref{dotq}), using $\dot m_{cs} = \dot q M/(1+q)^2$.

In the case of SQM, the inspiral continues to smaller separations
since $R_{SQM}<R_{NS}$.  As a consequence, mass transfer is
accelerated and the duration of the stable mass transfer outspiral
phase is shortened considerably.  In the SQM case the star barely
spirals out since $1/3-\alpha\simeq0$ (see Eq.~(\ref{dlnrdlnm2})).  The
star eventually loses all its mass. In contrast, for a neutron star
$1/3-\alpha>0$ and the star outspirals significantly, until
$\alpha\simeq-5/3$ when stable mass transfer terminates
(Eq.~(\ref{cond})).  This occurs above the minimum stable neutron star
mass, $\sim0.08$ M$_\odot$, so the remnant mini-neutron star will
resume spiralling in, concluding in a final merger.  In either case,
if stable mass transfer occurs, it proceeds for much longer than an
orbital period, perhaps up to a few tenths of a second.  The merger of
neutron star and self-bound star with a BH will significantly differ
in their gravity wave signatures (shown in the right panel of
Fig.~\ref{orbit}). The self-bound star case will have a shorter
timescale of emission, with probably higher peak luminosity, and will
lack a final increase due to resumption of inspiral in the NS case.

An additional contrast between mergers involving only NSs or only SQM stars
(but not BHs) is that the condition for stable mass transfer might not
be generally achievable for binary NSs.  Note that moderate mass NSs have
$\alpha\simeq0$, so Eq.~(\ref{cond}) requires $q<5/6$.  Had we used the
more exact formula of Eggleton, Eq. (\ref{eggleton}), we would have
found an even greater restriction $q<0.78$.   For example, the binary pulsar
PSR1913+16 has $q=0.96$.  In contrast, for small to moderate
mass self-bound SQM stars $\alpha \simeq 1/3$, so that stable mass
transfer can occur for $q<1$, which is its entire domain!

The evolution of self-bound SQM star - black hole merger including the
effects of non-conservative mass transfer, tidal synchronization,
the presence of an accretion disk, {\em etc.,} together with extensions to
include further effects of general relativity, will be reported separately.

In conclusion, compact star mergers offer a tantalizing possibility
for the detection via gravitational waves of strange quark matter in
the form of self-bound stars.

%%%%%%%%%%%%%%%%%%%%%%%%%%%%%%%%%%%%%%%%%%%%%%%%%%%%%%%%%%%%%%%%
\vspace*{-0.01in}
\begin{acknowledgments}
We thank Jes Madsen for helpful communications.  This work was
supported by the US-DOE grants DE-FG02-88ER40388 (for M.P.) and  
DE-FG02-87ER40317 (for J.M.L.).  
\end{acknowledgments}

\end{document}